\documentclass{article}
\usepackage{graphicx} 
\usepackage[table]{xcolor}
\usepackage{color, colortbl}

\usepackage{authblk}
\providecommand{\keywords}[1]{\textbf{\textit{Keywords:}} #1}

\usepackage{longtable}
\usepackage{booktabs}
\usepackage{array}
\usepackage{enumitem}

\usepackage{algorithm}
\usepackage{algpseudocode}
\usepackage{tikz}
\usetikzlibrary{trees,arrows.meta}
\usetikzlibrary{trees, positioning}
\usepackage{forest}
\usetikzlibrary{mindmap}

\title{Human-Oriented Image Retrieval System (HORSE): A Neuro-Symbolic Approach to Optimizing Retrieval of Previewed Images}
\author[1]{Abraham Itzhak Weinberg}
\affil[1]{AI-WEINBERG, AI Experts, Tel Aviv, Israel, aviw2010@gmail.com}

\begin{document}
\maketitle
\begin{abstract}
Image retrieval remains a challenging task due to the complex interaction between human visual perception, memory, and computational processes. Current image search engines often struggle to efficiently retrieve images based on natural language descriptions, as they rely on time-consuming preprocessing, tagging, and machine learning pipelines. This paper introduces the Human-Oriented Retrieval Search Engine for Images (HORSE), a novel approach that leverages neuro-symbolic indexing to improve image retrieval by focusing on human-oriented indexing. By integrating cognitive science insights with advanced computational techniques, HORSE enhances the retrieval process, making it more aligned with how humans perceive, store, and recall visual information. The neuro-symbolic framework combines the strengths of neural networks and symbolic reasoning, mitigating their individual limitations. The proposed system optimizes image retrieval, offering a more intuitive and efficient solution for users. We discuss the design and implementation of HORSE, highlight its potential applications in fields such as design error detection and knowledge management, and suggest future directions for research to further refine the system's metrics and capabilities.
\end{abstract}


\section{Introduction}
At the intersection of human expertise and artificial intelligence lies a fundamental challenge: how to effectively bridge the gap between human knowledge and computational intelligence. Traditional Symbolic AI approaches, which attempt to codify human expertise into rule-based systems, have long struggled with the closed world assumption problem—they can only reason within the boundaries of explicitly defined knowledge, limiting their adaptability to novel situations.
Conversely, modern Machine Learning (ML) approaches excel at pattern recognition but suffer from two critical limitations: biased outputs reflecting their training data and a lack of explainability that renders their decision-making processes opaque to human understanding.\\
This paper is organized as follows: first, we introduce the image retrieval tasks, their current solution approaches, and their drawbacks. Afterwards, we present the NeSY approach and explain how it can be integrated with image retrieval. Finally, we propose our HORSE algorithm and discuss it.\\

"Great living starts with a picture, held in your imagination, of what you would like to do or be" (Harry Emerson Fosdick, \cite{maltz2002new}). This concept of envisioning outcomes resonates with the Human-Oriented Image Retrieval System (HORSE), which aims to improve image retrieval by focusing on how humans mentally visualize and describe images. By leveraging neurosymbolic indexing, HORSE bridges the gap between human cognition and computational systems, enabling users to retrieve images based on natural language descriptions, much like imagining a desired outcome and bringing it into reality.\\
Our proposed HORSE algorithm offers a novel solution through a unique Neuro-Symbolic (NeSY) integration approach. Unlike conventional NeSY systems where reasoning processes are merely layered atop neural network outputs, HORSE begins by extracting relevant human knowledge and then implements AI processes specifically aligned with this knowledge foundation. This human-oriented integration ensures that computational reasoning remains compatible with human logic while leveraging the pattern recognition strengths of neural networks.
Retrieving visual information exemplifies the challenges this approach addresses. Current image retrieval systems face significant limitations in terms of human usability, often employing complex pipelines of preprocessing, tagging, and ML algorithms that create redundant feature storage while failing to optimize for previously viewed images. These systems frequently return results that, while algorithmically relevant, do not align with human perception and memory processes.\\
This research explores an alternative paradigm where users can retrieve previously viewed images by describing them in their natural language, creating a more intuitive human-computer interaction model. By optimizing retrieval processes around human memory and description capabilities, we aim to develop systems that work in harmony with human cognitive processes rather than requiring humans to adapt to machine limitations.
The research aims to achieve the following outcomes:
\begin{itemize}
\item A better understanding of the image retrieval process from both human and computer perspectives.
\item Insights into the interaction between human memory, storage, description, and retrieval.
\item  Development of human-computer metrics for evaluating the end-to-end retrieval process.
\item  An optimized image retrieval solution that improves accessibility and retrieval efficiency.
\end{itemize}

\subsection {Image Search Engines versus Image Retrieval and Indexing Engines}
The domain of visual information discovery and management is divided primarily between two types of systems, each serving distinct purposes and user needs. Image search engines, exemplified by Google Images \cite{fergus2004visual} and Bing Images \cite{hu2018web}, are designed for general users seeking to discover web-based images through text queries. These consumer-oriented platforms offer straightforward interfaces with basic filtering options for characteristics like size, color, and image type, prioritizing accessibility and breadth of coverage over technical sophistication.\\
In contrast, image retrieval and indexing engines serve technical and enterprise applications with specialized capabilities for managing curated image collections. Systems like Apache Solr with image extensions \cite{shahi2015apache} and Elasticsearch with image plugins \cite{dixit2016elasticsearch} support both text-based and content-based queries, enabling more precise access to visual assets. \\
These platforms incorporate advanced features including visual/image similarity matching\footnote{Image similarity can be thought of as a numerical representation of how alike two images are in terms of their visual content. There are several dimensions along which images can be similar, such as color, shape, texture, and composition.}\cite{santini1999similarity} or Structural Similarity Index (SSIM)\footnote{A commonly used metric that evaluates the structural similarity between two images. It takes into account luminance, contrast, and structure, providing a score ranging from -1 (completely dissimilar) to 1 (identical).}, automated feature extraction\footnote{It employs specialized algorithms or deep networks to automatically extract features from signals or images, eliminating the need for human intervention \cite{mckeown2000performance}.}, content classification\footnote{It classifies images into predefined categories based on their visual content \cite{park2004content}.}, and comprehensive metadata indexing—capabilities that support professional workflows in fields ranging from digital asset management to medical imaging.\\
The fundamental differences between these system types lie in their scope, query methodologies, and technical depth as can be seen in Table \ref{tab:image-search-comparison}. While search engines cast a wider net across the public web with primarily text-based queries, retrieval systems offer deeper analysis capabilities within defined collections using multiple query methods. This distinction reflects their different purposes: search engines connect users with previously unknown images, while retrieval systems help users efficiently locate and leverage known visual assets within managed repositories. The technical sophistication of retrieval systems comes with increased complexity but enables the precision and analytical capabilities required for professional and enterprise applications.
\begin{table}
\centering
\begin{tabular}{p{3cm} p{5cm} p{5cm}}
\toprule
\textbf{Characteristic} & \textbf{Image Search Engines} & \textbf{Image Retrieval and Indexing Engines} \\
\midrule
Target Users & General consumers & Technical professionals and enterprises \\
\midrule
Primary Purpose & Web-based image discovery & Management of curated image collections \\
\midrule
Examples & Google Images, Bing Images & Apache Solr with image extensions, Elasticsearch with image plugins \\
\midrule
Indexing Methodology & Web crawling with indexing based on surrounding text, file names, and basic visual features & Sophisticated multi-dimensional indexing structures (k-d trees, R-trees, locality-sensitive hashing) \\
\midrule
Query Types & Primarily text-based queries & Multiple query paradigms: text-based, query-by-example, query-by-sketch, hybrid approaches \\
\midrule
Query Processing & Keyword matching against indexed text & Complex similarity searches across feature vectors with semantic understanding \\
\midrule
Feature Extraction & Basic image features and metadata & Advanced visual feature extraction with domain-specific optimizations \\
\midrule
Data Sources & Public web content & Curated collections with structured metadata \\
\midrule
Data Management & Continuous discovery via web crawling with limited dataset control & Carefully maintained collections with version control and access permissions \\
\midrule
Analytical Capabilities & Basic filtering (size, color, type) & Advanced capabilities (object detection, scene understanding, facial recognition) \\
\midrule
Computational Focus & Speed and relevance across massive datasets & Precision within defined domains with domain-specific algorithms \\
\midrule
Integration Options & Limited API access & Robust APIs and integration frameworks for enterprise applications \\
\midrule
Customization & Generalized algorithms across all content & Can be tailored for domain-specific visual patterns and use cases \\
\midrule
Technical Complexity & Designed for ease of use & Higher complexity with more sophisticated controls \\
\midrule
Scope & Broader coverage across public web & Deeper analysis within defined collections \\
\midrule
Scalability Focus & Horizontal scaling for billions of images & Precise indexing and retrieval for domain-specific collections \\
\bottomrule
\end{tabular}
\caption{Comparison between Image Search Engines and Image Retrieval and Indexing Engines}
\label{tab:image-search-comparison}
\end{table}

\subsection{Background and Related Work}
Our proposed approach begins by examining the human image retrieval process. Human memory is capable of rapidly encoding visual information and storing it compactly for long-term retrieval \cite{wang2016context}, a characteristic that our approach seeks to leverage for efficient indexing and storage in computer systems. The novelty of this work lies in its integration of human cognitive factors into image retrieval, which, to the best of our knowledge, has not been widely explored.\\
HORSE translates these characteristics into NeSY rules. These rules are then used as human knowledge, enhancing further ML and retrieval steps. As far as we know, this integration has not been done before. Additionally, the translation of human memory into NeSY rules is unique in the field.\\
Our algorithm addresses the growing need for accessible image retrieval systems in the context of the increasing volume of images stored and shared on social media platforms and personal devices. The findings of this research will also have significant implications for improving image accessibility, knowledge management, and aiding professionals like designers and draftsmen in detecting design errors.\\

\subsection{Key Approaches and Algorithms in Image Retrieval and Indexing Systems}
Image retrieval and indexing systems rely on various approaches and algorithms that help efficiently manage and search visual data. One key area is feature extraction \cite{mckeown2000performance}, which involves extracting information from images in different forms. Color features, for example, are often used to capture the distribution of colors across an image, with techniques like color histograms, color moments, and color correlograms \cite{singh2012image}. Color histograms track the distribution of colors, while color moments capture the mean, variance, and skewness of color distributions, and the color correlogram represents the spatial correlation of colors within an image. Dominant color descriptors are also important for identifying key colors that characterize the overall appearance of an image.\\
Texture features focus on analyzing the surface properties of images \cite{armi2019texture}, with methods such as the Gray Level Co-occurrence Matrix (GLCM), which studies the spatial relationships between pixel intensities \cite{park2020measuring}. Gabor filters detect specific frequency content in various directions \cite{kamarainen2006invariance}, and Local Binary Patterns (LBP) capture local texture patterns \cite{pietikainen2015two}. Wavelet transforms offer a multi-resolution approach to texture analysis, which is crucial in many applications \cite{busch2004wavelet}.\\
Shape features are another critical aspect, and methods like edge detection, including Canny and Sobel operators, highlight the edges of an image \cite{jing2022recent}. Contour representations capture the shape outline \cite{yang2022overview}, while moment invariants provide a way to describe shapes in a rotation- and scale-invariant manner \cite{qi2021survey}. Shape contexts are also employed to capture the spatial relationships between points on the shape's boundary, offering a distinct representation of an object.\\
Deep learning-based approaches have gained significant attention, especially with Convolutional Neural Networks (CNNs), which are commonly used for feature extraction. Pre-trained models like Visual Geometry Group (VGG), Residual Network (ResNet), and Inception are often employed to generate feature vectors that serve as effective image descriptors \cite{gkelios2021deep}. Siamese networks have become popular for learning similarity metrics between image pairs, improving the ability to distinguish between similar and dissimilar images \cite{melekhov2016siamese}. Auto-encoders are applied for dimensionality reduction, simplifying the data representation while preserving key features. Self-supervised learning methods, which allow for better representations of images without requiring labeled data, have also proven beneficial in enhancing retrieval performance.\\
Indexing structures are essential for efficient search and retrieval. Tree-based methods, such as KD-trees (K-Dimensional) for low-dimensional features and R-trees for spatial indexing, are commonly used \cite{li2024survey}. M-trees are suitable for indexing in metric spaces \cite{chen2022indexing}, and VP-trees (Vantage-Point) are particularly effective for high-dimensional data \cite{ukey2023survey}. Hashing methods, including Locality Sensitive Hashing (LSH), Semantic Hashing, Spectral Hashing, and Product Quantization, also play a vital role in reducing the complexity of retrieval tasks, allowing for faster searches by encoding data into more compact representations \cite{han2024hash}.\\
Similarity measures are used to assess the closeness between images based on their features \cite{zhang2011fsim}. Euclidean distance is commonly used for comparing feature vectors, while cosine similarity is often employed in high-dimensional spaces. Earth Mover's Distance is useful for comparing histograms by measuring the cost of transforming one distribution into another \cite{tang2013earth}. Hamming distance is applied to binary features \cite{ionescu2004fuzzy}, and Mahalanobis distance is used when the data involves correlated features, providing a more accurate similarity measure in such cases \cite{akyilmaz2017similarity}.\\
Content-Based Image Retrieval (CBIR) is one of the most popular retrieval techniques, allowing systems to retrieve images based on their content \cite{hameed2021content}. Query by Example (QBE) and relevance feedback mechanisms are often used to refine searches, and multi-feature fusion strategies combine multiple types of features for improved retrieval accuracy \cite{mei2014multimedia}. Cross-modal retrieval, which includes techniques such as text-to-image search and image-to-text mapping, allows for retrieving images based on textual descriptions and vice versa \cite{cheng2021deep}. These techniques rely on joint embedding spaces that bridge the gap between different modalities.\\
Modern optimization approaches like Approximate Nearest Neighbor (ANN) search are widely used to improve retrieval efficiency \cite{tu2020approximate}. Tools like FAISS (Facebook AI Similarity Search), Annoy (Spotify's ANN library), and HNSW (Hierarchical Navigable Small World) enable faster nearest neighbor searches, reducing computational overhead in large datasets \cite{rahman2024optimizing}. When evaluating retrieval systems, metrics such as precision and recall, Mean Average Precision (MAP), and Normalized Discounted Cumulative Gain (NDCG) are commonly used to assess the relevance and ranking of results \cite{moffat2017computing}. Other important factors include retrieval time and memory efficiency, which are critical in real-world applications.\\
For real-world implementation, several considerations must be addressed. Scalability is crucial to handle large datasets, and update mechanisms are required to accommodate dynamic collections. Storage optimization ensures efficient data storage, while query optimization improves the speed of query processing. Load balancing in distributed systems is also necessary to handle varying demands and ensure smooth operation.\\
Enhanced retrieval methods, such as using multiple query images and query expansion, can help improve retrieval accuracy \cite{xie2014contextual}. Semantic search techniques improve retrieval by understanding the meaning behind search queries \cite{fernandez2011semantically}, while attribute-based filtering allows for more fine-grained searches based on specific image attributes \cite{hossen2024attribute}. Spatial verification is another technique that ensures the returned results align with the spatial characteristics of the query image \cite{zhou2013sift}.\\
Finally, advanced topics in image retrieval and indexing include fine-grained image retrieval, which focuses on retrieving images with subtle differences, and instance-level retrieval, which targets specific instances of objects. Cross-view image matching addresses challenges in matching images of the same object from different angles \cite{shi2020looking}, and zero-shot image retrieval enables retrieval of images without prior examples or training data \cite{kan2020zero}. Continuous learning systems, which adapt and improve over time through new data, are another emerging area in the field, ensuring that image retrieval systems remain effective as they evolve \cite{yang2023knowledge}.

\subsection{Notable Image Retrieval and Indexing Engines}
There are various notable image retrieval and indexing engines, which can be categorized into commercial, open source, and deep learning-based solutions. Commercial solutions include Elastic Image Search \cite{amato2018large}, which offers visual search capabilities within the Elasticsearch ecosystem, and Amazon Rekognition \cite{sharma2022object}, an AWS service designed for image analysis and similarity search. Google Cloud Vision Product Search enables visual product search and cataloging \cite{zhong2013real}, while Microsoft Azure Computer Vision provides image analysis and visual search capabilities \cite{meena2016architecture}.\\
Additionally, Algolia Visual Search serves as a visual search add-on for their search platform \cite{dasic2016applications}. Open-source solutions include Milvus, a distributed vector database that handles image feature vectors \cite{wang2021milvus}, and FAISS (Facebook AI Similarity Search), a high-performance similarity search library \cite{danopoulos2019approximate}. Qdrant is another vector similarity search engine that offers extended filtering support \cite{singh2023analyzing}, while Vearch is a high-performance vector similarity search engine developed by Jina AI \cite{sun2022product}. Vespa is a real-time big data serving engine that also supports image search \cite{singh2023analyzing}, and the ImageHash Python library allows simple image matching using perceptual hashes \cite{diyasa2020reverse}.\\
LIRE (Lucene Image REtrieval) is a Java library designed for image retrieval based on Lucene \cite{lux2013lire}. On the deep learning front, CLIP (Contrastive Language Image Pre-Training) is a model developed by OpenAI for text-to-image search \cite{gao2024clip}, and DupDetector uses neural networks to identify duplicate or similar images \cite{vanam2021identifying}. DeepSight is another deep learning-based visual search engine \cite{zhang2020deepsite}. These solutions typically offer several key features such as CBIR, reverse image search, near-duplicate detection, feature vector indexing, perceptual hashing, and multimodal search, which combines text and images, all with scalable distributed indexing capabilities. A comparison of these image search solutions and their technical approaches is provided in Table \ref{table:image_search_comparison}, which summarizes their key approaches and core algorithms.

\begin{table}
    \centering
    \begin{tabular}{p{4cm}p{6cm}p{5cm}}
    \toprule
    \textbf{Solution} & \textbf{Key Approaches} & \textbf{Core Algorithms} \\
    \midrule
    Elastic Image Search & Vector-based similarity search with CNN feature extraction & HNSW graphs\newline
    Approximate k-NN\newline
    Dense vector scoring\newline
    Cosine similarity computation \\
    \midrule
    Amazon Rekognition & Multi-model approach combining detection and recognition & Deep CNNs\newline
    Cascade classifiers\newline
    Siamese networks\newline
    YOLO variants \\
    \midrule
    Google Cloud Vision & AutoML-based custom model training with efficient feature extraction & EfficientNet backbone\newline
    Contrastive learning\newline
    Deep metric learning\newline
    BERT text-image matching \\
    \midrule
    Microsoft Azure Vision & Transformer-based architecture with multi-task capabilities & ResNet extraction\newline
    Scene graph generation\newline
    Faster R-CNN\newline
    Few-shot learning \\
    \midrule
    Milvus & Distributed vector search with multiple index support & IVF/HNSW/ANNOY indexing\newline
    GPU acceleration\newline
    Dynamic quantization\newline
    SIMD optimization \\
    \midrule
    FAISS & High-performance similarity search with compression & Product Quantization\newline
    Inverted File Index\newline
    Multi-probe LSH\newline
    Cluster-based indexing \\
    \midrule
    Qdrant & Graph-based vector search with filtering & HNSW indexing\newline
    Payload filtering\newline
    Segment storage\newline
    Query optimization \\
    \midrule
    CLIP & Contrastive learning between image and text & Vision Transformer\newline
    Zero-shot classification\newline
    Cross-modal attention\newline
    Temperature scaling \\
    \midrule
    ImageHash & Perceptual hash-based image matching & Average hashing\newline
    Difference hashing\newline
    Wavelet hashing\newline
    Color moment hashing \\
    \midrule
    LIRE & Traditional computer vision features & SIFT/SURF descriptors\newline
    Color/Edge detection\newline
    Gabor textures\newline
    Edge histograms \\
    \bottomrule
    \end{tabular}
    \caption{Comparison of Image Search Solutions and Their Technical Approaches}
    \label{table:image_search_comparison}
\end{table}

\subsubsection{Google's Various Image Search and Retrieval Technologies}
Google's innovations in image search and retrieval technologies play a pivotal role in reshaping how users interact with and access visual information online. 
Google has developed various advanced image search and retrieval technologies. Google Images, the standard consumer-facing search engine, supports reverse image search using advanced computer vision and ML techniques, and integrates with Google Lens technology \cite{al2020awjedni}. \\
Google Lens is a mobile-first visual search tool that can identify objects, text, and landmarks in real-time and provides shopping capabilities for visual product search, along with text extraction and translation features \cite{maurya2023application}.
It is available as a standalone app and is integrated into other Google products. For enterprise applications, Google Cloud Vision AI offers multiple services, including Product Search for building retail catalog search systems, Vision AI for general-purpose image analysis and classification, AutoML Vision for custom model training \cite{bisong2019google}, and the Video Intelligence API for analyzing video content \cite{kc2024integrating}.\\
Google's internal technologies include PlaNet, a model for geographic location estimation from images \cite{xu2018vision}, DELF (DEep Local Features) for landmark recognition \cite{noh2017large}, SwAV (Swapping Assignments between Views) for self-supervised learning in image understanding \cite{jaiswal2020survey}, and MUM (Multimodal Unified Model), which processes both text and images simultaneously \cite{fei2024vemo}.

\section{Neuro-Symbolic Approaches to Image Information Retrieval}
NeSY approaches to image information retrieval combine the strengths of neural networks and symbolic reasoning to enhance the effectiveness of retrieval systems \cite{dietz2023neuro}. Traditional information retrieval models typically rely on either neural networks or symbolic approaches, each with distinct advantages and limitations. Neural networks excel at processing high-dimensional data and recognizing patterns within visual content, while symbolic reasoning offers interpretability, logical consistency, and the integration of structured human knowledge. The integration of these complementary strengths in NeSY systems provides a promising framework to address the complex challenges of image retrieval, bridging the semantic gap between low-level image features and high-level human understanding.\\
NeSY approaches offer several key advantages for image retrieval systems \cite{bouneffouf2022survey}. One major benefit is the ability to maintain interpretability while leveraging the pattern recognition capabilities of neural networks \cite{weinberg2025neurosymbolic}. Unlike pure neural models, which often function as "black boxes," NeSY systems retain transparency through their symbolic components, making the reasoning behind their decisions more understandable. Additionally, these approaches can explicitly integrate domain knowledge, human expertise, and cognitive principles that are difficult to capture through data alone. This knowledge integration is particularly valuable in specialized fields where labeled data may be scarce or incomplete.\\
Moreover, the incorporation of symbolic reasoning allows NeSY systems to perform logical inferences about spatial relationships, object properties, and semantic contexts—key aspects of human-like image understanding. This ability to reason about images adds a layer of depth that pure neural systems cannot match. The inclusion of symbolic knowledge also reduces the dependency on massive datasets, making NeSY approaches more viable in scenarios where training data is limited. Finally, the symbolic component acts as a safeguard, providing robustness against the common failure modes seen in purely neural approaches and improving overall system consistency and reliability.\\
The unique challenges of image retrieval make NeSY approaches particularly well-suited for this task. Visual information is inherently hierarchical and relational, with meaning emerging not only from individual objects but also from their spatial arrangement, context, and relationships \cite{silva2014visualization}. This complexity aligns well with the capabilities of NeSY systems. One of the key challenges in image retrieval is the substantial gap between pixel-level representations and semantic understanding. While neural networks can efficiently process raw visual data, they often struggle to capture the abstract relationships and contextual knowledge that humans rely on when describing or searching for images. Symbolic components help bridge this gap by explicitly representing higher-level concepts, such as spatial relationships or object properties, which are crucial for understanding and interpreting images.\\
Moreover, natural language queries for image retrieval often involve imprecise, subjective, or context-dependent terms that require interpretation beyond simple keyword matching \cite{datta2008image}. NeSY systems are particularly adept at handling such queries by mapping linguistic descriptions to visual features through symbolic representations of concepts like "above," "larger than," or "similar to." This ability to handle nuanced language makes NeSY systems more effective in real-world applications, where queries are rarely straightforward.\\
Another reason NeSY approaches are well-suited for image retrieval is that human memory for images operates on multiple levels of abstraction—from broad impressions to specific details. This mirrors the complementary processing of neural and symbolic components in NeSY systems, making them more aligned with human cognitive processes \cite{alford2021neurosymbolic}. This alignment allows NeSY systems to create retrieval mechanisms that feel intuitive and natural to users, further enhancing the user experience.\\
Furthermore, NeSY systems can leverage prior knowledge about common object relationships and spatial configurations to make inferences about images \cite{khan2025survey}. This ability reduces the need for large amounts of training data, which is often required for pure neural approaches to learn these relationships. In specialized domains, where training data may be limited but domain knowledge is rich, this advantage becomes particularly important. By combining the strengths of neural and symbolic approaches, these systems offer a powerful solution for overcoming the challenges of image retrieval, making them an ideal choice for applications that require both high accuracy and interpretability.

\subsection{Key Disadvantages and Limitations of Traditional Major Approaches to Image Retrieval}
Although various image retrieval and indexing techniques offer promising solutions, each approach is accompanied by inherent limitations that impact their performance in different contexts. Table \ref{tab:image_retrieval_disadvantages} provides a comprehensive overview of the key disadvantages associated with the most widely used methods in this field.\\
Traditional feature extraction methods, such as color, texture, and shape features, exhibit several drawbacks. Color features, for instance, are highly sensitive to illumination changes and perform poorly with grayscale images \cite{singh2012image}. Texture features, while useful in many scenarios, are computationally expensive and sensitive to rotation and scaling. Additionally, shape features often struggle with occlusions and complex or deformable objects, and they require clean segmentation for optimal performance. These limitations underscore the challenges faced by traditional approaches in real-world applications.\\
In contrast, deep learning-based approaches, such as CNNs and Siamese networks, have shown significant promise in image retrieval tasks \cite{gkelios2021deep,melekhov2016siamese}. However, these methods come with their own set of challenges. CNNs, for example, require large amounts of training data and are computationally intensive, which can make them impractical in resource-limited environments. Moreover, both CNNs and Siamese networks suffer from poor interpretability, and the risk of overfitting is a constant concern. Additionally, these methods typically have high memory usage.\\
Indexing structures, such as tree-based methods and hashing techniques, also present significant disadvantages \cite{li2024survey,han2024hash}. Tree-based indexing approaches degrade in high-dimensional spaces and suffer from issues related to unbalanced trees and high memory overhead. Hashing methods, while efficient in some cases, are prone to information loss due to quantization and require complex handling of collisions. These issues can lead to poor retrieval performance, especially in dynamic or high-dimensional datasets.\\
Similarity measures, such as Euclidean distance and cosine similarity, are fundamental in many image retrieval systems but have limitations that hinder their effectiveness \cite{zhang2011fsim}. Euclidean distance, for example, is highly sensitive to outliers and assumes equal feature weights, which may not always hold true in real-world data. Similarly, cosine similarity fails to account for magnitude differences, which can be a critical factor in certain applications. Both methods also struggle with high-dimensional data and lack the ability to understand semantic relationships between features.\\
Finally, modern optimization techniques, such as ANN search and cross-modal retrieval, offer advanced capabilities but are not without their challenges \cite{tu2020approximate}. ANN search faces a trade-off between accuracy and speed, and it requires careful tuning of parameters to balance performance. Cross-modal retrieval, which involves the alignment of different types of data such as images and text, suffers from semantic gaps between modalities and challenges related to paired data requirements. Moreover, methods like fine-grained retrieval require extensive annotations and can be computationally expensive, which limits their practicality for large-scale applications.

\begin{longtable}{p{4cm}p{4cm}p{6cm}}
\caption{Popular Image Retrieval and Indexing Approaches and Disadvatanges\label{tab:image_retrieval_disadvantages}} \\

\toprule
\textbf{Approach} & \textbf{Category} & \textbf{Key Disadvantages} \\
\midrule
\endfirsthead

\multicolumn{3}{c}{Table \ref{tab:image_retrieval_disadvantages} continued} \\
\toprule
\textbf{Approach} & \textbf{Category} & \textbf{Key Disadvantages} \\
\midrule
\endhead

\midrule
\multicolumn{3}{r}{\emph{Continued on next page}} \\
\endfoot

\bottomrule
\endlastfoot

Color Features & Traditional Feature Extraction & 
\begin{itemize}[nosep]
    \item Sensitive to illumination changes
    \item Ignores spatial relationships
    \item Fails with similar color distributions
    \item Poor with grayscale images
    \item High storage overhead
\end{itemize} \\
\midrule

Texture Features & Traditional Feature Extraction & 
\begin{itemize}[nosep]
    \item Computationally expensive
    \item Sensitive to rotation/scale
    \item Limited for non-textured images
    \item Poor with viewpoint changes
    \item Requires multiple descriptors
\end{itemize} \\
\midrule

Shape Features & Traditional Feature Extraction & 
\begin{itemize}[nosep]
    \item Highly sensitive to occlusion
    \item Struggles with complex shapes
    \item Poor with deformable objects
    \item Computationally intensive
    \item Requires clean segmentation
\end{itemize} \\
\midrule

CNN Features & Deep Learning & 
\begin{itemize}[nosep]
    \item Large training data requirement
    \item High computational cost
    \item Poor interpretability
    \item Overfitting risks
    \item Heavy memory usage
\end{itemize} \\
\midrule

Siamese Networks & Deep Learning & 
\begin{itemize}[nosep]
    \item Complex training pair selection
    \item Training instability
    \item Limited multi-class handling
    \item Domain-specific retraining
    \item High memory requirements
\end{itemize} \\
\midrule

Tree-based Indexing & Indexing Structures & 
\begin{itemize}[nosep]
    \item Degrades in high dimensions
    \item Unbalanced tree issues
    \item High memory overhead
    \item Expensive updates
    \item Poor for dynamic data
\end{itemize} \\
\midrule

Hashing Methods & Indexing Structures & 
\begin{itemize}[nosep]
    \item Information loss from quantization
    \item Hash function dependency
    \item Complex collision handling
    \item Multiple table requirements
    \item Precision-recall trade-off
\end{itemize} \\
\midrule

Euclidean Distance & Similarity Measures & 
\begin{itemize}[nosep]
    \item Outlier sensitivity
    \item Equal feature weight assumption
    \item Poor in high dimensions
    \item Fixed-length feature requirement
    \item No semantic understanding
\end{itemize} \\
\midrule

Cosine Similarity & Similarity Measures & 
\begin{itemize}[nosep]
    \item Ignores magnitude differences
    \item Poor with sparse data
    \item High-dimension sensitivity
    \item Vector space limitation
    \item Simple relationship modeling
\end{itemize} \\
\midrule

Query by Example & CBIR & 
\begin{itemize}[nosep]
    \item Query quality dependency
    \item Semantic gap issues
    \item Feature representation limits
    \item No abstract query support
    \item Multiple example needs
\end{itemize} \\
\midrule

Relevance Feedback & CBIR & 
\begin{itemize}[nosep]
    \item Required user interaction
    \item Time-consuming process
    \item Convergence issues
    \item User fatigue
    \item Implementation complexity
\end{itemize} \\
\midrule

ANN Search & Modern Optimization & 
\begin{itemize}[nosep]
    \item Accuracy-speed trade-off
    \item Complex parameter tuning
    \item High memory overhead
    \item Missed relevant results
    \item Data distribution dependency
\end{itemize} \\
\midrule

Cross-Modal Retrieval & Advanced Methods & 
\begin{itemize}[nosep]
    \item Semantic gap between modalities
    \item Paired data requirement
    \item Vocabulary limitations
    \item Modal alignment issues
    \item Training complexity
\end{itemize} \\
\midrule

Fine-grained Retrieval & Advanced Methods & 
\begin{itemize}[nosep]
    \item Detailed annotation needs
    \item High computation costs
    \item Visual sensitivity
    \item Domain-specific features
    \item Limited generalization
\end{itemize}

\end{longtable}

\subsection{Neuro-Symbolic Integration}
Nesy approaches aim to overcome the limitations of both by integrating neural networks for pattern recognition with symbolic reasoning mechanisms for logic and interpretability. This integration results in a more robust system capable of reasoning about images and queries in a way that pure neural networks or symbolic systems alone cannot. 

\subsubsection{Neural Components}
Neural components in a NeSY image retrieval system primarily handle unstructured data, such as images, by learning representations that capture the underlying semantics and visual relationships. In this section, we will describe the components of Embedding Generation \cite{bhuyan2024graph}, Feature Learning \cite{gong2025neuro}, and Natural Language Understanding \cite{hamilton2024neuro}, which are essential for processing both visual and textual data in a neuro-symbolic image retrieval system. \\
These components work together to process both visual and textual data in a meaningful way. Embedding generation is the first key step, where neural networks transform raw image data into dense vector representations \cite{bhuyan2024graph}. CNNs and Vision Transformers (ViTs) are commonly used in this process, enabling the extraction of image features while capturing contextual relationships between objects. Following this, feature learning allows neural models to automatically discover relevant patterns from raw image data, creating hierarchical representations at various levels of abstraction \cite{gong2025neuro}. This adaptability helps the system recognize domain-specific features. Finally, natural language understanding plays a crucial role in processing text queries, enabling the system to interpret natural language inputs, resolve ambiguity, and effectively match visual content with textual descriptions \cite{hamilton2024neuro}. Through these interconnected processes, the system is able to bridge the gap between visual and textual data, providing a robust and efficient retrieval mechanism.

\subsubsection{Symbolic Components}
Symbolic components in a NeSY system add a layer of reasoning and structure, ensuring that the retrieved information is logically coherent and interpretable. This section will describe the components of Knowledge Representation \cite{di2023towards}, Reasoning Mechanisms \cite{prentzas2021exploring}, and Constraint Management \cite{liu2024knowledge}.\\
Knowledge Representation involves the use of formal ontologies and taxonomies to define concepts such as “dog,” “car,” and “building,” along with their relationships \cite{di2023towards}. Symbolic representations, such as scene graphs and spatial relationships such as "above," "below," "contains", help to provide a deeper understanding of the image context. Reasoning Mechanisms utilize methods like forward and backward chaining to apply logical rules over the learned representations, clarifying relationships between different objects or concepts within an image and enhancing the retrieval process \cite{prentzas2021exploring}. Finally, Constraint Management ensures that the retrieved images meet specific criteria by enforcing logical constraints, such as matching certain attributes or fitting within predefined categories, thus ensuring the relevance and accuracy of the retrieval results \cite{liu2024knowledge}.

\subsubsection{Integration Mechanisms}
The integration of neural and symbolic components is a key aspect of NeSY systems, enabling the combination of learning-based and reasoning-based approaches for enhanced performance. 
This section will describe the following integration mechanisms:
Neural-to-Symbolic Translation \cite{pulicharla2025neurosymbolic}, Symbolic-to-Neural Guidance \cite{pulicharla2025neurosymbolic}, and Hybrid Reasoning Paths \cite{confalonieri2025multiple}.\\
One way integration occurs is through Neural-to-Symbolic Translation, where learned neural representations are transformed into symbolic forms that can be reasoned with. Visual features are mapped to symbols that are understood within the context of a predefined ontology \cite{pulicharla2025neurosymbolic}. Another important mechanism is Symbolic-to-Neural Guidance, where symbolic knowledge, such as logical rules, influences and refines the neural learning process. Constraints from the symbolic layer can guide the neural network outputs, ensuring the generation of more accurate or contextually appropriate features \cite{pulicharla2025neurosymbolic}. Additionally, Hybrid Reasoning Paths combine neural pattern recognition and symbolic inference in a synergistic manner. In this approach, the neural component is responsible for initial image feature extraction, while symbolic reasoning applies logical rules to filter, refine, and enhance the results, providing a more comprehensive retrieval mechanism.  

\subsection{Applications to Image Information Retrieval}
In the context of image retrieval, NeSY approaches offer several key advantages. These systems excel in handling complex queries that require both pattern recognition and logical reasoning \cite{bouneffouf2022survey}. For example, they can process queries like "find images with a red car," which involves pattern recognition, as well as more complex requests like "find images where the car is in front of a building," which necessitate logical reasoning. \\
Additionally, symbolic reasoning enables semantic search, where images are retrieved not only based on visual similarity but also by considering logical constraints and the semantic relationships between objects. NeSY systems also support multi-modal search, allowing users to query with a combination of text and images. For instance, users might upload a reference image and ask the system to find similar images, considering both visual features and the semantic meanings behind them. \\ Finally, the symbolic layer enhances explainability by providing interpretable results, allowing users to understand why certain images were retrieved based on explicit logical rules and relationships, rather than relying solely on the opaque decision-making processes of deep learning models.

\subsection{Example Architecture for Image Retrieval}
A typical architecture for a NeSY image retrieval system consists of several stages that work together to process and interpret both visual and textual data \cite{oltramari2020neuro}. The first stage, image processing, involves CNNs or ViTs, which process the image data to extract important features such as object recognition, localization, and contextual understanding. \\ 
Following this, the query parsing stage employs a Natural Language Processing (NLP) module to interpret user queries, converting text input into symbolic representations. For example, a query like "Find images with a red ball" would be mapped to a search for red-colored objects within images. Once the query is parsed, symbolic filtering is applied. In this stage, symbolic reasoning methods are used to refine the retrieved images based on logical constraints.\\
For instance, a constraint such as "only images where the ball is on a table" would narrow down the results. Finally, in the retrieval stage, both the neural and symbolic components collaborate to rank and return the most relevant images, considering not only visual similarity but also the logical constraints provided by the symbolic reasoning.

\section{Human-Oriented Image Retrieval System (HORSE) Methodology and Problem Definition}
As mentioned earlier, NeSY starts from human knowledge. Since the image retrieval target is to serve human users it has to be designed to their usage profile. We can diagnose the way that human approach for image retrieval. The transformation from retrieving image by using key words to a free language increase the challenge. \\
Analyzing the description of images shows that the human memory plays a main role in this task. As far as we researched, the human memory retrieval of images consists of several characteristics: image objects spacial relations, their size and characteristics such as color and if there are human also body and facial gesture.\\
Human memory plays a crucial role in image analysis and description, as research shows that when humans recall and describe images, they typically encode and retrieve various aspects of visual information. In this paper, we define several meta-rules based on human cognition and map them to the NeSY. These include spatial relationships between objects, often referred to as the "where" information, which helps to understand the positioning of elements within an image \cite{rosman2011learning,hollingworth2001see}.\\
Additionally, we consider object properties like size, color, and shape, which contribute to the detailed recognition of individual items \cite{hu2023effects,greene2022effects}. Finally, the semantic meaning and contextual relationships of objects within the image play an essential role in comprehending the broader narrative and interpretation of the visual content.

\begin{algorithm}
\caption{Human-Oriented Image Retrieval (HORSE)}
\label{alg:image_retrieval}
\begin{algorithmic}[1]
\State \textbf{Input:} Image dataset
\State \textbf{Output:} Matching images based on query
\State
\State \textbf{Step 1:} Extracting Human Retrieval Patterns
\State \textbf{Step 2:} Translate the Retrieval Patterns into NeuroSymbolic Meta Rules
\State
\For{each image}
    \State \textbf{Step 3.1:} Detect Meaningful Objects (using OCR, for instance)
    \State \textbf{Step 3.2:} Extract properties of each Meaningful Object
    \State \textbf{Step 3.3:} According to Step 2, find relations between Meaningful Objects
\EndFor
\State
\textbf{Step 4:} Index the images using their meaningful objects and relations
\State
\State \textbf{Step 5:} Enable NLP search query
\State \textbf{Step 5.1:} Extract the query objects and relations
\State \textbf{Step 5.2:} Search the image indexed database for matching images
\State
\State \textbf{Output:} View the matching images
\end{algorithmic}
\end{algorithm}

Human memory plays a crucial role in image analysis and description, as research shows that when humans recall and describe images, they typically encode and retrieve various aspects of visual information. This process is often hierarchical and gist-based, with people typically remembering the overall meaning or impression of an image first, followed by specific details. Additionally, emotionally significant elements tend to be remembered more vividly, highlighting the role of emotional salience in memory retrieval \cite{cooper2019memories,baker2021searching}.\\
Moreover, memory retrieval for images involves both bottom-up (feature-driven) and top-down (knowledge/expectation-driven) processes working together, rather than relying on feature extraction alone \cite{ochsner2009bottom,shelton2017dynamic}. HORSE algorithm proposes the NeSY approach, which combines insights from human memory with AI techniques for image retrieval. The proposed approach exemplifies the extraction of rules from human knowledge, supported by psychological insights and the human memory retrieval system. These rules can be based on organizational data and generalized beyond past data, incorporating research on human memory, decision-making, and brain characteristics.\\
The HORSE algorithm \ref{alg:image_retrieval} follows several steps: recognizing objects in the image and their names, as well as human emotions; extracting object characteristics like color; normalizing object sizes and ranking them by relative size; mapping relationships between objects using 2D and 3D relations (such as A being above B, or to the left/right, or behind/in front); and learning relations based on an image corpus. For instance, the system learns that, in 99\% of cases, a car is on the ground, the sky is in the upper part of images, and a house is bigger than a human. The system can then identify the uniqueness of an image by comparing it to other images and a 'normal' baseline. To develop a solution that bridges the human visual system and memory with computational systems, it is essential to take an interdisciplinary approach. Traditional algorithms often fail to account for human cognitive processes, which can limit the reliability and efficiency of image retrieval. Thus, our methodology examines the image retrieval process from both the human and computer perspectives. \\
Key parameters for evaluating this solution include human-side factors such as memory functionality, psychological aspects, and linguistic description capabilities, as well as computer-side factors like algorithm complexity, running time, and storage compactness. By assessing the retrieval process from these angles, we aim to provide a solution that optimally balances human needs and computational efficiency. 

\section{Discussion}
The proposed HORSE algorithm represents a significant advancement in the field of image retrieval by addressing fundamental limitations of traditional approaches. By incorporating NeSY principles that mirror human cognitive processes, HORSE offers several advantages that warrant further discussion.

\subsection{Alignment with Human Cognitive Processes}
Our approach deliberately mirrors the hierarchical and relational nature of human visual memory. Traditional computer vision systems often emphasize raw feature extraction or pure statistical learning, which can create a mismatch between how machines index images and how humans naturally recall them. HORSE bridges this gap by establishing a framework that captures spatial relationships, object properties, and semantic meanings—the three key dimensions that characterize human image recall as identified in our research.\\
The incorporation of both bottom-up (feature-driven) and top-down (knowledge-based) processes in our algorithm acknowledges the bidirectional nature of human image processing \cite{ochsner2009bottom,shelton2017dynamic}. This dual-process approach allows the system to balance concrete visual features with contextual understanding, making it particularly effective for natural language queries that may contain imprecise or subjective descriptions.

\subsection{Technical Implications and Advantages}
The NeSY foundation of HORSE offers several technical advantages over purely neural or purely symbolic approaches. By extracting meta-rules from human cognitive patterns, the system can operate with greater interpretability than black-box deep learning models, while maintaining more flexibility than rigid rule-based systems. This middle-ground approach is particularly valuable for debugging, system refinement, and trustworthiness.\\
Our indexing strategy based on meaningful objects and their relationships represents a more efficient computational approach than exhaustive feature extraction. By focusing on elements that would be salient to human memory, we potentially reduce the dimensionality of the search space without sacrificing retrieval accuracy. This efficiency becomes increasingly important as image databases grow in size.\\
The normalization of object sizes and relative positioning in three-dimensional space allows HORSE to generalize across images with different perspectives and scales. This capability addresses a common limitation in traditional image retrieval systems, which often struggle with viewpoint invariance.

\subsection{Limitations and Future Work}
Despite its advantages, HORSE faces several challenges that require further research. First, the extraction of accurate spatial relationships depends on reliable object detection and scene understanding, which remain active research areas. Errors in object recognition can propagate through the system, potentially affecting retrieval accuracy.
Second, while our approach aims to mimic human memory patterns, individual differences in visual perception and memory remain a challenge. Future iterations of HORSE could benefit from personalization mechanisms that learn individual users' recall patterns and adjust accordingly.\\
Third, the current implementation primarily focuses on static images. Extending the framework to video retrieval would require additional considerations for temporal relationships and motion patterns, which are crucial aspects of human memory for dynamic visual content.
Future work should address these limitations while exploring several promising directions:

\begin{enumerate}
    \item {Multimodal Integration}: Incorporating audio descriptions, text captions, and other contextual metadata could enhance retrieval accuracy, especially for ambiguous queries.
    
    \item {Adaptive Learning}: Developing mechanisms for HORSE to continuously refine its understanding of human memory patterns based on user interactions and feedback.
    
    \item {Cross-Cultural Validation}: Testing the system across diverse cultural contexts to ensure that the extracted meta-rules generalize across different user populations.
    
    \item {Computational Optimization}: Further refining the indexing structures to balance comprehensiveness with computational efficiency, particularly for large-scale image collections.
\end{enumerate}

\subsection{Broader Applications}
The principles underlying HORSE extend beyond simple image retrieval. The same NeSY approach could be applied to several adjacent domains:

\begin{itemize}
    \item {Visual Anomaly Detection}: By establishing normative relationships between objects (e.g., "cars are typically on roads"), the system could identify unusual or incorrect images.
    
    \item {Accessibility Tools}: HORSE could facilitate image descriptions for visually impaired users by focusing on the aspects of images that sighted humans find most memorable.
    
    \item {Educational Applications}: The system could support visual learning by helping students locate relevant images based on conceptual descriptions rather than keywords alone.
    
    \item {Design Assistance}: Creative professionals could use natural language descriptions to retrieve inspirational images that match their conceptual vision.
\end{itemize}

\section{Conclusion}
In this paper, we proposed a novel approach for human-oriented image retrieval, utilizing neuro-symbolic indexing. The method takes into account the user's cognitive and linguistic abilities, in addition to standard computational parameters, to optimize the image retrieval process. Future work will focus on refining the metrics and exploring potential applications in areas like design error detection and knowledge management.\\
HORSE represents a promising step toward more human-centric image retrieval systems. By grounding computational approaches in cognitive science research, we aim to create systems that feel more intuitive and accessible to users. The NeSY framework offers a balanced approach that maintains the advantages of both neural networks and symbolic reasoning while mitigating their respective limitations.\\
As visual content continues to proliferate across digital platforms, the need for effective retrieval systems becomes increasingly critical. HORSE demonstrates that by better understanding how humans process, store, and recall visual information, we can design more effective computational systems that serve human needs. Future research should continue exploring this intersection of cognitive science and computer vision to further enhance human-computer interaction in visual domains.


\bibliographystyle{IEEEtran}
\bibliography{ref.bib}

\begin{thebibliography}{10}
\providecommand{\url}[1]{#1}
\csname url@samestyle\endcsname
\providecommand{\newblock}{\relax}
\providecommand{\bibinfo}[2]{#2}
\providecommand{\BIBentrySTDinterwordspacing}{\spaceskip=0pt\relax}
\providecommand{\BIBentryALTinterwordstretchfactor}{4}
\providecommand{\BIBentryALTinterwordspacing}{\spaceskip=\fontdimen2\font plus
\BIBentryALTinterwordstretchfactor\fontdimen3\font minus \fontdimen4\font\relax}
\providecommand{\BIBforeignlanguage}[2]{{%
\expandafter\ifx\csname l@#1\endcsname\relax
\typeout{** WARNING: IEEEtran.bst: No hyphenation pattern has been}%
\typeout{** loaded for the language `#1'. Using the pattern for}%
\typeout{** the default language instead.}%
\else
\language=\csname l@#1\endcsname
\fi
#2}}
\providecommand{\BIBdecl}{\relax}
\BIBdecl

\bibitem{maltz2002new}
M.~Maltz, \emph{New psycho-cybernetics}.\hskip 1em plus 0.5em minus 0.4em\relax Penguin, 2002.

\bibitem{fergus2004visual}
R.~Fergus, P.~Perona, and A.~Zisserman, ``A visual category filter for google images,'' in \emph{Computer Vision-ECCV 2004: 8th European Conference on Computer Vision, Prague, Czech Republic, May 11-14, 2004. Proceedings, Part I 8}.\hskip 1em plus 0.5em minus 0.4em\relax Springer, 2004, pp. 242--256.

\bibitem{hu2018web}
H.~Hu, Y.~Wang, L.~Yang, P.~Komlev, L.~Huang, X.~Chen, J.~Huang, Y.~Wu, M.~Merchant, and A.~Sacheti, ``Web-scale responsive visual search at bing,'' in \emph{Proceedings of the 24th ACM SIGKDD international conference on knowledge discovery \& data mining}, 2018, pp. 359--367.

\bibitem{shahi2015apache}
D.~Shahi, \emph{Apache Solr: a practical approach to enterprise search}.\hskip 1em plus 0.5em minus 0.4em\relax Springer, 2015.

\bibitem{dixit2016elasticsearch}
B.~Dixit, \emph{Elasticsearch essentials}.\hskip 1em plus 0.5em minus 0.4em\relax Packt Publishing Ltd, 2016.

\bibitem{santini1999similarity}
S.~Santini and R.~Jain, ``Similarity measures,'' \emph{IEEE Transactions on pattern analysis and machine Intelligence}, vol.~21, no.~9, pp. 871--883, 1999.

\bibitem{mckeown2000performance}
D.~M. McKeown, T.~Bulwinkle, S.~Cochran, W.~Harvey, C.~McGlone, and J.~A. Shufelt, ``Performance evaluation for automatic feature extraction,'' \emph{International Archives of Photogrammetry and Remote Sensing}, vol.~33, no. B2; PART 2, pp. 379--394, 2000.

\bibitem{park2004content}
S.~B. Park, J.~W. Lee, and S.~K. Kim, ``Content-based image classification using a neural network,'' \emph{Pattern Recognition Letters}, vol.~25, no.~3, pp. 287--300, 2004.

\bibitem{wang2016context}
L.~Wang, X.~Zhao, Y.~Si, L.~Cao, and Y.~Liu, ``Context-associative hierarchical memory model for human activity recognition and prediction,'' \emph{IEEE Transactions on Multimedia}, vol.~19, no.~3, pp. 646--659, 2016.

\bibitem{singh2012image}
S.~M. Singh and K.~Hemachandran, ``Image retrieval based on the combination of color histogram and color moment,'' \emph{International journal of computer applications}, vol.~58, no.~3, 2012.

\bibitem{armi2019texture}
L.~Armi and S.~Fekri-Ershad, ``Texture image analysis and texture classification methods-a review,'' \emph{arXiv preprint arXiv:1904.06554}, 2019.

\bibitem{park2020measuring}
Y.~Park and J.-M. Guldmann, ``Measuring continuous landscape patterns with gray-level co-occurrence matrix (glcm) indices: An alternative to patch metrics?'' \emph{Ecological Indicators}, vol. 109, p. 105802, 2020.

\bibitem{kamarainen2006invariance}
J.-K. Kamarainen, V.~Kyrki, and H.~Kalviainen, ``Invariance properties of gabor filter-based features-overview and applications,'' \emph{IEEE Transactions on image processing}, vol.~15, no.~5, pp. 1088--1099, 2006.

\bibitem{pietikainen2015two}
M.~Pietik{\"a}inen and G.~Zhao, ``Two decades of local binary patterns: A survey,'' in \emph{Advances in independent component analysis and learning machines}.\hskip 1em plus 0.5em minus 0.4em\relax Elsevier, 2015, pp. 175--210.

\bibitem{busch2004wavelet}
A.~W. Busch, ``Wavelet transform for texture analysis with application to document analysis,'' Ph.D. dissertation, Queensland University of Technology, 2004.

\bibitem{jing2022recent}
J.~Jing, S.~Liu, G.~Wang, W.~Zhang, and C.~Sun, ``Recent advances on image edge detection: A comprehensive review,'' \emph{Neurocomputing}, vol. 503, pp. 259--271, 2022.

\bibitem{yang2022overview}
D.~Yang, B.~Peng, Z.~Al-Huda, A.~Malik, and D.~Zhai, ``An overview of edge and object contour detection,'' \emph{Neurocomputing}, vol. 488, pp. 470--493, 2022.

\bibitem{qi2021survey}
S.~Qi, Y.~Zhang, C.~Wang, J.~Zhou, and X.~Cao, ``A survey of orthogonal moments for image representation: Theory, implementation, and evaluation,'' \emph{ACM Computing Surveys (CSUR)}, vol.~55, no.~1, pp. 1--35, 2021.

\bibitem{gkelios2021deep}
S.~Gkelios, A.~Sophokleous, S.~Plakias, Y.~Boutalis, and S.~A. Chatzichristofis, ``Deep convolutional features for image retrieval,'' \emph{Expert Systems with Applications}, vol. 177, p. 114940, 2021.

\bibitem{melekhov2016siamese}
I.~Melekhov, J.~Kannala, and E.~Rahtu, ``Siamese network features for image matching,'' in \emph{2016 23rd international conference on pattern recognition (ICPR)}.\hskip 1em plus 0.5em minus 0.4em\relax IEEE, 2016, pp. 378--383.

\bibitem{li2024survey}
M.~Li, H.~Wang, H.~Dai, M.~Li, C.~Chai, R.~Gu, F.~Chen, Z.~Chen, S.~Li, Q.~Liu \emph{et~al.}, ``A survey of multi-dimensional indexes: past and future trends,'' \emph{IEEE Transactions on Knowledge and Data Engineering}, vol.~36, no.~8, pp. 3635--3655, 2024.

\bibitem{chen2022indexing}
L.~Chen, Y.~Gao, X.~Song, Z.~Li, Y.~Zhu, X.~Miao, and C.~S. Jensen, ``Indexing metric spaces for exact similarity search,'' \emph{ACM Computing Surveys}, vol.~55, no.~6, pp. 1--39, 2022.

\bibitem{ukey2023survey}
N.~Ukey, Z.~Yang, B.~Li, G.~Zhang, Y.~Hu, and W.~Zhang, ``Survey on exact knn queries over high-dimensional data space,'' \emph{Sensors}, vol.~23, no.~2, p. 629, 2023.

\bibitem{han2024hash}
L.~Han, M.~E. Paoletti, X.~Tao, Z.~Wu, J.~M. Haut, P.~Li, R.~Pastor-Vargas, and A.~Plaza, ``Hash-based remote sensing image retrieval,'' \emph{IEEE Transactions on Geoscience and Remote Sensing}, 2024.

\bibitem{zhang2011fsim}
L.~Zhang, L.~Zhang, X.~Mou, and D.~Zhang, ``Fsim: A feature similarity index for image quality assessment,'' \emph{IEEE transactions on Image Processing}, vol.~20, no.~8, pp. 2378--2386, 2011.

\bibitem{tang2013earth}
Y.~Tang, L.~H. U, Y.~Cai, N.~Mamoulis, and R.~Cheng, ``Earth mover's distance based similarity search at scale,'' \emph{Proceedings of the VLDB Endowment}, vol.~7, no.~4, pp. 313--324, 2013.

\bibitem{ionescu2004fuzzy}
M.~Ionescu and A.~Ralescu, ``Fuzzy hamming distance in a content-based image retrieval system,'' in \emph{2004 IEEE International Conference on Fuzzy Systems (IEEE Cat. No. 04CH37542)}, vol.~3.\hskip 1em plus 0.5em minus 0.4em\relax IEEE, 2004, pp. 1721--1726.

\bibitem{akyilmaz2017similarity}
E.~Akyilmaz and U.~M. Leloglu, ``Similarity ratio based adaptive mahalanobis distance algorithm to generate sar superpixels,'' \emph{Canadian Journal of Remote Sensing}, vol.~43, no.~6, pp. 569--581, 2017.

\bibitem{hameed2021content}
I.~M. Hameed, S.~H. Abdulhussain, and B.~M. Mahmmod, ``Content-based image retrieval: A review of recent trends,'' \emph{Cogent Engineering}, vol.~8, no.~1, p. 1927469, 2021.

\bibitem{mei2014multimedia}
T.~Mei, Y.~Rui, S.~Li, and Q.~Tian, ``Multimedia search reranking: A literature survey,'' \emph{ACM Computing Surveys (CSUR)}, vol.~46, no.~3, pp. 1--38, 2014.

\bibitem{cheng2021deep}
Q.~Cheng, Y.~Zhou, P.~Fu, Y.~Xu, and L.~Zhang, ``A deep semantic alignment network for the cross-modal image-text retrieval in remote sensing,'' \emph{IEEE Journal of Selected Topics in Applied Earth Observations and Remote Sensing}, vol.~14, pp. 4284--4297, 2021.

\bibitem{tu2020approximate}
Z.~Tu, W.~Yang, Z.~Fu, Y.~Xie, L.~Tan, K.~Xiong, M.~Li, and J.~Lin, ``Approximate nearest neighbor search and lightweight dense vector reranking in multi-stage retrieval architectures,'' in \emph{Proceedings of the 2020 ACM SIGIR on International Conference on Theory of Information Retrieval}, 2020, pp. 97--100.

\bibitem{rahman2024optimizing}
M.~Rahman, S.~M.~E. Rabbi, and M.~M. Rashid, ``Optimizing domain-specific image retrieval: A benchmark of faiss and annoy with fine-tuned features,'' \emph{arXiv preprint arXiv:2412.01555}, 2024.

\bibitem{moffat2017computing}
A.~Moffat, ``Computing maximized effectiveness distance for recall-based metrics,'' \emph{IEEE Transactions on Knowledge and Data Engineering}, vol.~30, no.~1, pp. 198--203, 2017.

\bibitem{xie2014contextual}
H.~Xie, Y.~Zhang, J.~Tan, L.~Guo, and J.~Li, ``Contextual query expansion for image retrieval,'' \emph{IEEE Transactions on Multimedia}, vol.~16, no.~4, pp. 1104--1114, 2014.

\bibitem{fernandez2011semantically}
M.~Fern{\'a}ndez, I.~Cantador, V.~L{\'o}pez, D.~Vallet, P.~Castells, and E.~Motta, ``Semantically enhanced information retrieval: An ontology-based approach,'' \emph{Journal of Web Semantics}, vol.~9, no.~4, pp. 434--452, 2011.

\bibitem{hossen2024attribute}
M.~B. Hossen, Z.~Ye, A.~Abdussalam, and S.~U. Hassan, ``Attribute-driven filtering: A new attributes predicting approach for fine-grained image captioning,'' \emph{Engineering Applications of Artificial Intelligence}, vol. 137, p. 109134, 2024.

\bibitem{zhou2013sift}
W.~Zhou, H.~Li, Y.~Lu, and Q.~Tian, ``Sift match verification by geometric coding for large-scale partial-duplicate web image search,'' \emph{ACM Transactions on Multimedia Computing, Communications, and Applications (TOMM)}, vol.~9, no.~1, pp. 1--18, 2013.

\bibitem{shi2020looking}
Y.~Shi, X.~Yu, D.~Campbell, and H.~Li, ``Where am i looking at? joint location and orientation estimation by cross-view matching,'' in \emph{Proceedings of the IEEE/CVF Conference on Computer Vision and Pattern Recognition}, 2020, pp. 4064--4072.

\bibitem{kan2020zero}
S.~Kan, Y.~Cen, Y.~Cen, M.~Vladimir, Y.~Li, and Z.~He, ``Zero-shot learning to index on semantic trees for scalable image retrieval,'' \emph{IEEE Transactions on Image Processing}, vol.~30, pp. 501--516, 2020.

\bibitem{yang2023knowledge}
R.~Yang, S.~Wang, H.~Zhang, S.~Xu, Y.~Guo, X.~Ye, B.~Hou, and L.~Jiao, ``Knowledge decomposition and replay: A novel cross-modal image-text retrieval continual learning method,'' in \emph{Proceedings of the 31st ACM International Conference on Multimedia}, 2023, pp. 6510--6519.

\bibitem{amato2018large}
G.~Amato, P.~Bolettieri, F.~Carrara, F.~Falchi, and C.~Gennaro, ``Large-scale image retrieval with elasticsearch,'' in \emph{The 41st International ACM SIGIR Conference on Research \& Development in Information Retrieval}, 2018, pp. 925--928.

\bibitem{sharma2022object}
V.~Sharma, ``Object detection and recognition using amazon rekognition with boto3,'' in \emph{2022 6th International Conference on Trends in Electronics and Informatics (ICOEI)}.\hskip 1em plus 0.5em minus 0.4em\relax IEEE, 2022, pp. 727--732.

\bibitem{zhong2013real}
Y.~Zhong, P.~J. Garrigues, and J.~P. Bigham, ``Real time object scanning using a mobile phone and cloud-based visual search engine,'' in \emph{Proceedings of the 15th International ACM SIGACCESS Conference on Computers and Accessibility}, 2013, pp. 1--8.

\bibitem{meena2016architecture}
M.~Meena, A.~R. Singh, and V.~A. Bharadi, ``Architecture for software as a service (saas) model of cbir on hybrid cloud of microsoft azure,'' \emph{Procedia Computer Science}, vol.~79, pp. 569--578, 2016.

\bibitem{dasic2016applications}
P.~Dasic, J.~Dasic, and J.~Crvenkovic, ``Applications of the search as a service (saas),'' \emph{Bulletin of the Transilvania University of Brasov. Series I-Engineering Sciences}, pp. 91--98, 2016.

\bibitem{wang2021milvus}
J.~Wang, X.~Yi, R.~Guo, H.~Jin, P.~Xu, S.~Li, X.~Wang, X.~Guo, C.~Li, X.~Xu \emph{et~al.}, ``Milvus: A purpose-built vector data management system,'' in \emph{Proceedings of the 2021 International Conference on Management of Data}, 2021, pp. 2614--2627.

\bibitem{danopoulos2019approximate}
D.~Danopoulos, C.~Kachris, and D.~Soudris, ``Approximate similarity search with faiss framework using fpgas on the cloud,'' in \emph{International Conference on Embedded Computer Systems}.\hskip 1em plus 0.5em minus 0.4em\relax Springer, 2019, pp. 373--386.

\bibitem{singh2023analyzing}
P.~N. Singh, S.~Talasila, and S.~V. Banakar, ``Analyzing embedding models for embedding vectors in vector databases,'' in \emph{2023 IEEE International Conference on ICT in Business Industry \& Government (ICTBIG)}.\hskip 1em plus 0.5em minus 0.4em\relax IEEE, 2023, pp. 1--7.

\bibitem{sun2022product}
C.~Sun, L.~Bin~Song, and L.~Ying, ``Product re-identification system in fully automated defect detection,'' in \emph{International Conference on Smart Multimedia}.\hskip 1em plus 0.5em minus 0.4em\relax Springer, 2022, pp. 144--156.

\bibitem{diyasa2020reverse}
I.~G. S.~M. Diyasa, A.~D. Alhajir, A.~M. Hakim, and M.~F. Rohman, ``Reverse image search analysis based on pre-trained convolutional neural network model,'' in \emph{2020 6th Information Technology International Seminar (ITIS)}.\hskip 1em plus 0.5em minus 0.4em\relax IEEE, 2020, pp. 1--6.

\bibitem{lux2013lire}
M.~Lux, ``Lire: Open source image retrieval in java,'' in \emph{Proceedings of the 21st ACM international conference on Multimedia}, 2013, pp. 843--846.

\bibitem{gao2024clip}
P.~Gao, S.~Geng, R.~Zhang, T.~Ma, R.~Fang, Y.~Zhang, H.~Li, and Y.~Qiao, ``Clip-adapter: Better vision-language models with feature adapters,'' \emph{International Journal of Computer Vision}, vol. 132, no.~2, pp. 581--595, 2024.

\bibitem{vanam2021identifying}
D.~Vanam and V.~R. Pulipati, ``Identifying duplicate questions in community question answering forums using machine learning approaches,'' in \emph{Machine Learning Technologies and Applications: Proceedings of ICACECS 2020}.\hskip 1em plus 0.5em minus 0.4em\relax Springer, 2021, pp. 131--140.

\bibitem{zhang2020deepsite}
Y.~Zhang, S.~Qiao, S.~Ji, and Y.~Li, ``Deepsite: bidirectional lstm and cnn models for predicting dna--protein binding,'' \emph{International Journal of Machine Learning and Cybernetics}, vol.~11, pp. 841--851, 2020.

\bibitem{al2020awjedni}
H.~Al-Lohibi, T.~Alkhamisi, M.~Assagran, A.~Aljohani, and A.~O. Aljahdali, ``Awjedni: a reverse-image-search application,'' \emph{ADCAIJ: Advances in Distributed Computing and Artificial Intelligence Journal}, vol.~9, no.~3, p.~49, 2020.

\bibitem{maurya2023application}
A.~Maurya, A.~Kumar, and D.~Alimohammadi, ``Application of google lens to promote information services beyond the traditional techniques,'' \emph{Qualitative and Quantitative Methods in Libraries}, vol.~12, no.~1, pp. 111--136, 2023.

\bibitem{bisong2019google}
E.~Bisong, ``Google automl: cloud vision,'' in \emph{Building Machine Learning and Deep Learning Models on Google Cloud Platform: A Comprehensive Guide for Beginners}.\hskip 1em plus 0.5em minus 0.4em\relax Springer, 2019, pp. 581--598.

\bibitem{kc2024integrating}
A.~KC and S.~Aravind, ``Integrating google ai: Enhancing library services for the digital age,'' 2024.

\bibitem{xu2018vision}
Y.~Xu, L.~Pan, C.~Du, J.~Li, N.~Jing, and J.~Wu, ``Vision-based uavs aerial image localization: A survey,'' in \emph{Proceedings of the 2nd ACM SIGSPATIAL International Workshop on AI for Geographic Knowledge Discovery}, 2018, pp. 9--18.

\bibitem{noh2017large}
H.~Noh, A.~Araujo, J.~Sim, T.~Weyand, and B.~Han, ``Large-scale image retrieval with attentive deep local features,'' in \emph{Proceedings of the IEEE international conference on computer vision}, 2017, pp. 3456--3465.

\bibitem{jaiswal2020survey}
A.~Jaiswal, A.~R. Babu, M.~Z. Zadeh, D.~Banerjee, and F.~Makedon, ``A survey on contrastive self-supervised learning,'' \emph{Technologies}, vol.~9, no.~1, p.~2, 2020.

\bibitem{fei2024vemo}
N.~Fei, H.~Jiang, H.~Lu, J.~Long, Y.~Dai, T.~Fan, Z.~Cao, and Z.~Lu, ``Vemo: A versatile elastic multi-modal model for search-oriented multi-task learning,'' in \emph{European Conference on Information Retrieval}.\hskip 1em plus 0.5em minus 0.4em\relax Springer, 2024, pp. 56--72.

\bibitem{dietz2023neuro}
L.~Dietz, H.~Bast, S.~Chatterjee, J.~Dalton, J.-Y. Nie, and R.~Nogueira, ``Neuro-symbolic representations for information retrieval,'' in \emph{Proceedings of the 46th International ACM SIGIR Conference on Research and Development in Information Retrieval}, 2023, pp. 3436--3439.

\bibitem{bouneffouf2022survey}
D.~Bouneffouf and C.~C. Aggarwal, ``Survey on applications of neurosymbolic artificial intelligence,'' \emph{arXiv preprint arXiv:2209.12618}, 2022.

\bibitem{weinberg2025neurosymbolic}
A.~I. Weinberg, ``Neurosymbolic ai and mechanistic interpretability: Can they align in the artificial general intelligence era?'' 2025.

\bibitem{silva2014visualization}
I.~C. S.~d. Silva, ``Visualization of intensional and extensional levels of ontologies,'' 2014.

\bibitem{datta2008image}
R.~Datta, D.~Joshi, J.~Li, and J.~Z. Wang, ``Image retrieval: Ideas, influences, and trends of the new age,'' \emph{ACM Computing Surveys (Csur)}, vol.~40, no.~2, pp. 1--60, 2008.

\bibitem{alford2021neurosymbolic}
S.~Alford, ``A neurosymbolic approach to abstraction and reasoning,'' Ph.D. dissertation, Massachusetts Institute of Technology, 2021.

\bibitem{khan2025survey}
M.~J. Khan, F.~Ilievski, J.~G. Breslin, and E.~Curry, ``A survey of neurosymbolic visual reasoning with scene graphs and common sense knowledge,'' \emph{Neurosymbolic Artificial Intelligence}, vol.~1, pp. NAI--240\,719, 2025.

\bibitem{bhuyan2024graph}
B.~P. Bhuyan, A.~Ramdane-Cherif, T.~P. Singh, and R.~Tomar, ``Graph neural networks in neural-symbolic computing,'' in \emph{Neuro-Symbolic Artificial Intelligence: Bridging Logic and Learning}.\hskip 1em plus 0.5em minus 0.4em\relax Springer, 2024, pp. 231--253.

\bibitem{gong2025neuro}
N.~Gong, W.~Ying, D.~Wang, and Y.~Fu, ``Neuro-symbolic embedding for short and effective feature selection via autoregressive generation,'' \emph{ACM Transactions on Intelligent Systems and Technology}, vol.~16, no.~2, pp. 1--21, 2025.

\bibitem{hamilton2024neuro}
K.~Hamilton, A.~Nayak, B.~Bo{\v{z}}i{\'c}, and L.~Longo, ``Is neuro-symbolic ai meeting its promises in natural language processing? a structured review,'' \emph{Semantic Web}, vol.~15, no.~4, pp. 1265--1306, 2024.

\bibitem{di2023towards}
P.~Di~Maio, ``Towards a web standard for neuro-symbolic integration and knowledge representation using model cards,'' in \emph{Data Science with Semantic Technologies}.\hskip 1em plus 0.5em minus 0.4em\relax CRC Press, 2023, pp. 173--193.

\bibitem{prentzas2021exploring}
J.~Prentzas and I.~Hatzilygeroudis, ``Exploring aspects regarding reasoning in neuro-symbolic rules and connectionist expert systems,'' in \emph{2021 12th International Conference on Information, Intelligence, Systems \& Applications (IISA)}.\hskip 1em plus 0.5em minus 0.4em\relax IEEE, 2021, pp. 1--8.

\bibitem{liu2024knowledge}
J.~Liu, ``Knowledge completion method based on interference principle,'' in \emph{2024 6th International Conference on Frontier Technologies of Information and Computer (ICFTIC)}.\hskip 1em plus 0.5em minus 0.4em\relax IEEE, 2024, pp. 398--401.

\bibitem{pulicharla2025neurosymbolic}
M.~R. Pulicharla, ``Neurosymbolic ai: Bridging neural networks and symbolic reasoning,'' 2025.

\bibitem{confalonieri2025multiple}
R.~Confalonieri and G.~Guizzardi, ``On the multiple roles of ontologies in explanations for neuro-symbolic ai,'' \emph{Neurosymbolic Artificial Intelligence}, vol.~1, pp. NAI--240\,754, 2025.

\bibitem{oltramari2020neuro}
A.~Oltramari, J.~Francis, C.~Henson, K.~Ma, and R.~Wickramarachchi, ``Neuro-symbolic architectures for context understanding,'' in \emph{Knowledge Graphs for eXplainable Artificial Intelligence: Foundations, Applications and Challenges}.\hskip 1em plus 0.5em minus 0.4em\relax IOS Press, 2020, pp. 143--160.

\bibitem{rosman2011learning}
B.~Rosman and S.~Ramamoorthy, ``Learning spatial relationships between objects,'' \emph{The International Journal of Robotics Research}, vol.~30, no.~11, pp. 1328--1342, 2011.

\bibitem{hollingworth2001see}
A.~Hollingworth, C.~C. Williams, and J.~M. Henderson, ``To see and remember: Visually specific information is retained in memory from previously attended objects in natural scenes,'' \emph{Psychonomic Bulletin \& Review}, vol.~8, no.~4, pp. 761--768, 2001.

\bibitem{hu2023effects}
Z.~Hu and J.~Yang, ``Effects of memory cue and interest in remembering and forgetting of gist and details,'' \emph{Frontiers in Psychology}, vol.~14, p. 1244288, 2023.

\bibitem{greene2022effects}
N.~R. Greene and M.~Naveh-Benjamin, ``The effects of divided attention at encoding on specific and gist-based associative episodic memory,'' \emph{Memory \& Cognition}, vol.~50, no.~1, pp. 59--76, 2022.

\bibitem{cooper2019memories}
R.~A. Cooper, E.~A. Kensinger, and M.~Ritchey, ``Memories fade: The relationship between memory vividness and remembered visual salience,'' \emph{Psychological Science}, vol.~30, no.~5, pp. 657--668, 2019.

\bibitem{baker2021searching}
A.~L. Baker, M.~Kim, and J.~E. Hoffman, ``Searching for emotional salience,'' \emph{Cognition}, vol. 214, p. 104730, 2021.

\bibitem{ochsner2009bottom}
K.~N. Ochsner, R.~R. Ray, B.~Hughes, K.~McRae, J.~C. Cooper, J.~Weber, J.~D. Gabrieli, and J.~J. Gross, ``Bottom-up and top-down processes in emotion generation: common and distinct neural mechanisms,'' \emph{Psychological science}, vol.~20, no.~11, pp. 1322--1331, 2009.

\bibitem{shelton2017dynamic}
J.~T. Shelton and M.~K. Scullin, ``The dynamic interplay between bottom-up and top-down processes supporting prospective remembering,'' \emph{Current Directions in Psychological Science}, vol.~26, no.~4, pp. 352--358, 2017.

\end{thebibliography}

\end{document}